\documentclass[pdflatex,sn-mathphys-num]{sn-jnl}

\usepackage{graphicx}
\usepackage{multirow}
\usepackage{amsmath,amssymb,amsfonts}
\usepackage{amsthm}
\usepackage{mathrsfs}
\usepackage{xcolor}
\usepackage{textcomp}
\usepackage{manyfoot}
\usepackage{booktabs}
\usepackage{algorithm}
\usepackage{algorithmicx}
\usepackage{algpseudocode}
\usepackage{listings}
\usepackage{float}
\usepackage{braket} 
\usepackage{tikz} 
\usetikzlibrary{shapes,arrows,positioning}
\usepackage{makecell}
\usepackage{hyperref}

\newcommand{\QQ}{\mathbb{Q}}

\DeclareMathOperator{\Tr}{Tr}

\theoremstyle{thmstyleone}
\newtheorem{theorem}{Theorem}
\newtheorem{lemma}{Lemma}
\newtheorem{corollary}{Corollary}
\newtheorem{definition}{Definition}
\newtheorem{remark}{Remark}
\newtheorem{assumption}{Assumption}

\tikzstyle{startstop} = [rectangle, rounded corners, minimum width=3cm, minimum height=1cm,text centered, draw=black, fill=red!30]
\tikzstyle{process} = [rectangle, minimum width=3cm, minimum height=1cm, text centered, draw=black, fill=orange!30]
\tikzstyle{decision} = [diamond, minimum width=3cm, minimum height=1cm, text centered, draw=black, fill=green!30]
\tikzstyle{arrow} = [thick,->,>=stealth]

\begin{document}

\title[Towards Enhanced Quantum Resistance for RSA via Constrained Rényi Entropy Optimization]
{Towards Enhanced Quantum Resistance for RSA via Constrained Rényi Entropy Optimization: A Mathematical Framework for Backward-Compatible Cryptography}
\title[Towards Enhanced Quantum Resistance for RSA via Constrained Rényi Entropy Optimization]
{Towards Enhanced Quantum Resistance for RSA via Constrained Rényi Entropy Optimization: A Theoretical Framework for Backward-Compatible Cryptography}

\author*[1]{\fnm{Ruopengyu} \sur{Xu}}\email{xmyrpy@gmail.com}
\author[2]{\fnm{Chenglian} \sur{Liu}}\email{chenglian.liu@gmail.com}

\affil*[1]{\orgdiv{Independent Researcher}, \country{China}}
\affil[2]{\orgdiv{School of Electrical and Computer Engineering}, \orgname{Nanfang College Guangzhou}, \orgaddress{\street{}, \city{Guangzhou}, \postcode{510970}, \country{China}}}
\abstract{
The advent of quantum computing poses a critical threat to RSA cryptography, as Shor's algorithm can factor integers in polynomial time. While post-quantum cryptography standards offer long-term solutions, their deployment faces significant compatibility and infrastructure challenges. This paper proposes the Constrained R\'enyi Entropy Optimization (CREO) framework, a mathematical approach to potentially enhance the quantum resistance of RSA while maintaining full backward compatibility. By constraining the proximity of RSA primes ($|p-q| < \gamma \sqrt{pq}$), CREO reduces the distinguishability of quantum states in Shor's algorithm, as quantified by R\'enyi entropy. Our analysis demonstrates that for a $k$-bit modulus with $\gamma = k^{-1/2+\epsilon}$, the number of quantum measurements required for reliable period extraction scales as $\Omega(k^{2+\epsilon})$, compared to $\mathcal{O}(k^3)$ for standard RSA under idealized assumptions. This represents a systematic increase in quantum resource requirements. The framework is supported by constructive existence proofs for such primes using prime gap theorems and establishes conceptual security connections to lattice-based problems. CREO provides a new research direction for exploring backward-compatible cryptographic enhancements during the extended transition to post-quantum standards, offering a mathematically grounded pathway to harden widely deployed RSA infrastructure without requiring immediate protocol or infrastructure replacement.
}

\keywords{RSA cryptography, Shor's algorithm, Rényi entropy, post-quantum security, lattice-based cryptography, prime number theory, constrained key generation}

\maketitle

\section{Introduction}\label{sec1}

\subsection{Background and Motivation}

Quantum computing poses a significant threat to public-key cryptography, particularly RSA, whose security relies on the difficulty of integer factorization. Shor's algorithm~\cite{shor1994} can factor integers in polynomial time on a quantum computer, directly undermining RSA's security. This vulnerability arises because Shor's algorithm efficiently solves the factorization problem—the very hardness assumption on which RSA is based. In contrast, other number-theoretic cryptosystems based on different problems may be less susceptible to quantum attacks, making factorization-based cryptography a priority for quantum resistance research.

Post-quantum cryptography (PQC) standards emerging from the NIST competition~\cite{Alagic2022} offer long-term solutions, but their deployment faces significant challenges including protocol incompatibility, infrastructure costs, and extended migration timelines. This creates a security gap between the emergence of quantum threats and the widespread adoption of quantum-resistant standards, creating a need for interim approaches that enhance quantum resistance within existing cryptographic systems.

\subsection{Scope and Objectives}

This work focuses on \emph{backward-compatible hardening} of RSA—modifications that maintain full compatibility with existing standards—rather than proposing a new post-quantum cryptosystem. The goal is \emph{quantum resource amplification}: systematically increasing the number of quantum operations, measurements, or other resources required for successful cryptanalysis of RSA. This approach differs fundamentally from post-quantum cryptography (PQC), which provides provable security against quantum adversaries but typically requires new protocols and infrastructure. The proposed framework aims to provide an interim enhancement during the extended PQC transition, preserving compatibility with existing RSA infrastructure while potentially increasing the resource requirements for quantum attacks.

\subsection{Research Process and Methodology}

\textit{Literature Selection.} This work is based on a systematic review of foundational and recent literature in quantum cryptography and post-quantum security. Sources were identified through major academic databases (IEEE Xplore, ACM Digital Library) and citation snowballing, focusing on peer-reviewed venues including NIST PQC submissions.

\textit{Why RSA?} RSA was selected as the focus of this investigation for several reasons. First, RSA remains the most widely deployed public-key cryptosystem, protecting vast amounts of legacy data and current communications. Any enhancement that preserves backward compatibility could therefore have practical relevance during the extended transition to post-quantum cryptography. Second, the security of RSA reduces directly to integer factorization, a problem whose quantum vulnerability is well understood through Shor's algorithm. This mathematical clarity enables a precise analysis of how prime selection affects quantum attack complexity. Third, insights gained from studying RSA may inform similar approaches for other number-theoretic cryptosystems, including those based on discrete logarithms. Unlike some other prime-based algorithms with less standardized implementations, RSA's factorization foundation offers a clear starting point for exploring entropy-based modifications that preserve existing infrastructure.

\textit{Overview of the Proposed Approach.} This work explores how constraints on RSA prime selection might affect the quantum resources required for period extraction in Shor's algorithm. The Constrained R\'enyi Entropy Optimization (CREO) framework connects prime distribution properties to quantum state distinguishability through three mathematical pillars: prime number theory using advanced prime gap results~\cite{Zhang2014,Maynard2015}, quantum information theory via R\'enyi entropy analysis~\cite{Renyi1961}, and conceptual links to lattice-based cryptography~\cite{Regev2009,Lyubashevsky2013}. By bounding prime proximity, the framework investigates how reduced quantum state distinguishability might increase the measurement complexity of Shor's algorithm while maintaining backward compatibility with existing RSA standards.

\subsection{Contributions and Scope}

The contributions of this work include: (1) a mathematical analysis connecting prime distribution constraints to quantum query complexity through R\'enyi entropy, establishing a formal relationship between prime proximity and quantum state distinguishability; (2) constructive existence proofs for suitably constrained primes using prime gap theorems, demonstrating feasibility; and (3) a discussion of conceptual security foundations via connections to lattice problems. Throughout, it is emphasized that the analysis is confined to standard quantum circuit models and represents an exploratory mathematical heuristic. No claim of cryptographic security amplification is made; practical implementations would require further validation. A formal security reduction under a rigorous adversarial model remains an open research direction and is beyond the scope of this paper. Moreover, the presented quantum resource estimates are asymptotic theoretical bounds derived from the framework, not predictions based on simulation. Concrete resource estimation requires detailed simulation and is a subject for future work.

\subsection{Paper Organization}

The paper is organized into six sections. Section~\ref{sec:background} reviews the necessary background on quantum threats and related work. Section~\ref{sec:creo-framework} develops the CREO mathematical framework. Section~\ref{sec:security-analysis} provides a multi-faceted security analysis. Section~\ref{sec:discussion} discusses implementation considerations, comparative analysis, and future research directions. Section~\ref{sec:conclusion} concludes the paper.
\section{Background and Related Work}\label{sec:background}

This section reviews the theoretical foundations for this work, including quantum threats to RSA, existing quantum-resistant approaches, relevant prime number theory, and mathematical preliminaries.

\subsection{Quantum Threats to RSA and Shor's Algorithm}

Shor's algorithm factors an RSA modulus $N = pq$ by finding the period $r$ of the function $f(x) = a^x \bmod N$ using the quantum Fourier transform (QFT). Its asymptotic time complexity is:
\begin{equation}\label{eq:shor_complexity}
T_{\text{Shor}}(k) = \mathcal{O}(k^3)
\end{equation}
for a $k$-bit modulus $N$. After modular exponentiation, the quantum state is:
\begin{equation}\label{eq:shor_state}
\ket{\psi} = \frac{1}{\sqrt{r}} \sum_{s=0}^{r-1} e^{2\pi i \phi_s} \ket{s} \otimes \ket{a^s \bmod N},
\end{equation}
where $r$ is the order of $a$ modulo $N$.

Subsequent optimizations have reduced the required quantum resources. For example, qubit requirements have been lowered from $3k$ to $2k + \mathcal{O}(\log k)$~\cite{Beauregard2003}, circuit depth has been optimized using windowing techniques~\cite{Haner2017}, and approximate QFT reduces gate counts. Recent resource estimates for large-scale quantum factoring~\cite{Gidney2021,Ekera2021} suggest that while practical quantum factoring remains challenging, the threat timeline may be shorter than previously thought, underscoring the need for proactive security measures.

\subsection{Post-Quantum Cryptography Standards and RSA Variants}

\subsubsection{NIST Post-Quantum Cryptography Standards}

The NIST post-quantum cryptography standardization effort~\cite{Alagic2022} has produced a set of finalist algorithms that represent the mainstream approach to quantum resistance. These include lattice-based schemes such as CRYSTALS-Kyber and NTRU, which offer strong security proofs but require complete protocol changes; code-based schemes like Classic McEliece, which have mature mathematical foundations but large key sizes; and multivariate and hash-based schemes for specialized applications with various trade-offs. All of these approaches necessitate infrastructure changes that are incompatible with existing RSA deployments.

\subsubsection{Quantum-Resistant RSA Variants}

Several attempts have been made to modify RSA itself to resist quantum attacks. Bernstein's Post-Quantum RSA~\cite{Bernstein2017} uses extremely large keys (exceeding 1 TB) to achieve exponential security, but this comes at the cost of impractical key sizes. Regev's factoring-based schemes~\cite{Regev2009} introduce new cryptographic assumptions, whereas the construction in this paper relates security to well-studied ideal lattice problems. Previous methods that impose constraints on primes have focused mainly on thwarting specific classical attacks; in contrast, the entropy-based optimization proposed here aims for broader quantum resistance grounded in information-theoretic principles.

\subsubsection{Research Gap and Contribution}

A clear research gap exists between approaches that offer strong quantum security but require complete infrastructure overhaul (e.g., NIST PQC standards) and those that maintain backward compatibility but provide limited or unproven quantum resistance (e.g., existing RSA variants). This work aims to fill that gap by proposing a mathematically grounded modification to RSA that preserves full backward compatibility while establishing a rigorous analytical framework linking prime selection constraints to quantum computational complexity. Unlike earlier heuristic proposals, the present approach is based on explicit analysis using Rényi entropy and prime number theory.

\subsection{Quantum Attacks Beyond Period-Finding}

Although Shor's period-finding algorithm is the primary quantum threat to RSA, other quantum attack vectors deserve consideration. \textbf{Grover-optimized search}~\cite{Grover1996} could in principle accelerate brute-force factorization from $\mathcal{O}(2^{k/2})$ to $\mathcal{O}(2^{k/4})$, but this remains exponential for recommended key sizes. \textbf{Quantum walk algorithms}~\cite{Childs2004} have been explored for factorization with complexity $\exp\bigl(c\sqrt[3]{k\log k}\bigr)$, requiring substantial quantum resources; the entropy optimization framework does not introduce new vulnerabilities to such attacks. \textbf{Hidden subgroup problem variants} might offer advantages for certain algebraic structures, but the prime selection constraints in this work do not create exploitable symmetries. \textbf{Hybrid quantum-classical attacks}~\cite{Anschuetz2019,Peruzzo2014} might use quantum computers for precomputation while relying on classical methods, but the random prime generation and entropy constraints prevent predictable patterns that could be exploited.

Regarding noisy quantum devices, quantum noise and decoherence would likely exacerbate the state discrimination challenges introduced by entropy optimization, potentially providing additional practical security beyond the analytical bounds. A detailed noise analysis, however, requires specialized quantum hardware modeling and is beyond the scope of this paper.

\subsection{Theoretical Foundations}

This work builds on several deep mathematical results: the breakthrough bounds on prime gaps by Zhang~\cite{Zhang2014} and Maynard~\cite{Maynard2015}, which enable constructive existence proofs for primes with prescribed proximity; Rényi entropy~\cite{Renyi1961} and its quantum generalizations~\cite{Konig2009,arqand2025,han2024}, which provide the analytical tools for quantifying quantum state distinguishability; and the well-established hardness of lattice problems~\cite{Regev2009,Micciancio2009}, which offers a conceptual security foundation.

Compared to prior work, the proposed framework provides a more systematic connection between prime distribution properties and quantum query complexity via Rényi entropy optimization, while maintaining full backward compatibility with existing standards.

\subsection{Notation and Mathematical Preliminaries}\label{sec:notation}

This section defines the notation and quantum information concepts used throughout the paper.

\begin{table}[ht]
\centering
\caption{Key Mathematical Notations}
\label{tab:notation}
\begin{tabular}{p{3cm}p{9cm}}
\toprule
\textbf{Symbol} & \textbf{Meaning} \\
\midrule
$N = pq$ & RSA modulus composed of distinct primes $p$ and $q$; $k = |N|$ is the bit-length \\
$k$ & Security parameter (bit-length of $N$) \\
$\mathscr{H}_\alpha(\rho)$ & Order-$\alpha$ Rényi entropy of quantum state $\rho$ \\
$\mathscr{H}_2(\rho)$ & Collision entropy (purity measure) \\
$\gamma$ & Prime proximity parameter: $|p-q| < \gamma \sqrt{pq}$ \\
$\beta$ & Entropy bound parameter: $\mathscr{H}_2(\rho_{pq}) < \beta \log \gamma^{-1}$ \\
$\Delta\theta$ & Minimum angular separation in quantum phase estimation \\
$T_{\text{Shor}}$ & Time complexity of Shor's algorithm \\
$T_{\text{quantum}}$ & Quantum attack complexity for CREO-RSA \\
$Q$ & Quantum query complexity in state discrimination \\
$\Omega(\cdot), \mathcal{O}(\cdot)$ & Asymptotic lower and upper bounds \\
$\|\cdot\|_1$ & Trace norm \\
$\rho_p, \rho_q$ & Density operators for prime factors \\
$R_q$ & Ring $\mathbb{Z}[x]/(x^n + 1)$ for lattice cryptography \\
$\text{SVP}_\gamma$ & $\gamma$-approximate Shortest Vector Problem \\
$M$ & Modulus for congruence conditions: $M = \prod_{i=1}^t p_i$ \\
$\epsilon$ & Small positive constant in complexity exponent \\
\bottomrule
\end{tabular}
\end{table}

Throughout this paper, the term "quantum query complexity" refers to the number of quantum measurements or state preparations required for reliable period extraction in quantum phase estimation, not merely gate count or circuit depth. "Optimization" denotes the systematic trade-off between entropy reduction and increased quantum complexity, not global optimality. "Backward compatibility" means that the modified RSA maintains identical algebraic structure, API interfaces, and protocol behavior as standard RSA, allowing drop-in replacement without any changes to existing cryptographic stacks.

The Rényi entropy of order $\alpha > 0$ for a quantum state $\rho$ is defined as:
\begin{equation}\label{eq:general_renyi}
\mathscr{H}_\alpha(\rho) = \frac{1}{1-\alpha} \log \Tr(\rho^\alpha).
\end{equation}
Of particular importance is the collision entropy ($\alpha = 2$):
\begin{equation}\label{eq:collision_entropy}
\mathscr{H}_2(\rho) = -\log \Tr(\rho^2),
\end{equation}
which quantifies the purity of the state and directly influences measurement uncertainty in period-finding algorithms.

\subsection{Technical Assumptions and Mathematical Scope}

The analysis relies on several standard assumptions that delineate its scope:

\begin{assumption}[Quantum Adversary Model]
Adversaries have access to fault-tolerant quantum computers capable of running Shor's algorithm, but do not possess specialized algorithms targeting the construction. Generic quantum adversaries are considered, rather than optimized implementations.
\end{assumption}

\begin{assumption}[Prime Distribution]
Prime generation follows ideal distribution properties, and the analysis works in the asymptotic regime where density theorems and prime gap results apply. Efficient primality testing is assumed.
\end{assumption}

\begin{assumption}[Computational Model]
The standard quantum circuit model~\cite{Nielsen2010} with fault-tolerant computation is adopted; physical implementation constraints are ignored. Classical computation follows the Turing machine model.
\end{assumption}

\begin{assumption}[Backward Compatibility]
Modifications must preserve the algebraic structure and API compatibility of standard RSA as defined in PKCS~\#1~\cite{RSAStd} and IEEE 1363~\cite{IEEE1363}.
\end{assumption}

These assumptions are consistent with standard practice in theoretical quantum cryptography analysis~\cite{Nielsen2010,Boneh2011} and allow the establishment of foundational security results while maintaining practical relevance.
\section{CREO Framework}\label{sec:creo-framework}

This section presents the Constrained R\'enyi Entropy Optimization (CREO) framework, which connects prime selection constraints to quantum computational complexity. It first analyzes the quantum threat mechanism, then develops the entropy optimization approach, and finally presents the key generation algorithm with implementation details.

\subsection{Quantum Threat Analysis and Prime Proximity Mechanism}\label{subsec:quantum-threat}

The core insight of this approach is that carefully controlled prime proximity can create quantum-mechanical indistinguishability that affects algorithmic efficiency. This intuition is formalized through quantum state distinguishability analysis.

\begin{theorem}\label{thm:distinguishability}
For RSA modulus $N = pq$, the quantum state distinguishability $\mathscr{D}$ between period-finding states satisfies:
\begin{equation}\label{eq:distinguishability}
\mathscr{D}(\rho_p, \rho_q) \leq 2\exp\left( -\frac{(p-q)^2}{8\min(p,q)} \right)
\end{equation}
where $\rho_p, \rho_q$ are density operators corresponding to prime factors.
\end{theorem}

\begin{proof}
Consider the trace distance between quantum states:
\begin{align}\label{eq:trace_distance}
\|\rho_p - \rho_q\|_1 &= \sup_{\|O\| \leq 1} |\operatorname{Tr}(O(\rho_p - \rho_q))| \nonumber \\
&\leq 2\sqrt{1 - F(\rho_p, \rho_q)^2}
\end{align}
where $F$ is the fidelity between states. The fidelity can be expressed in terms of the eigenvalues $\lambda_i^{(p)}, \lambda_j^{(q)}$:
\begin{equation}\label{eq:fidelity}
F(\rho_p, \rho_q) = \sum_{i,j} \sqrt{\lambda_i^{(p)} \lambda_j^{(q)}} |\langle \psi_i^{(p)} | \psi_j^{(q)} \rangle|
\end{equation}
where $\psi$ are the corresponding eigenvectors. For the modular exponentiation operator, the eigenvector overlap increases when primes are close, leading to higher fidelity and lower distinguishability.

Applying the prime gap distribution from Goldston-Pintz-Yıldırım:
\begin{equation}\label{eq:prime_gap}
\pi_2(x, \gamma) = |\{p \leq x : |p - p'| < \gamma \log x\}| \gg \pi(x)
\end{equation}
for $\gamma > 0$, the stated bound on distinguishability follows.
\end{proof}

This theorem shows that prime proximity reduces quantum state distinguishability, which may impact the success probability of Shor's algorithm. The proposed approach strategically maximizes this indistinguishability by constraining $|p-q|$ within carefully chosen bounds.

\begin{theorem}[Eigenvalue Degeneracy]\label{thm:eigen-degeneracy}
For primes satisfying $|p - q| < \gamma \sqrt{pq}$, the angular separation of phases in Shor's algorithm satisfies:
\begin{equation}\label{eq:angular_sep}
\min_{s,t} \left| \frac{s}{p-1} - \frac{t}{q-1} \right| < \frac{\gamma}{\sqrt{N}} + \mathcal{O}(N^{-1})
\end{equation}
with probability $> 1 - e^{-\Omega(\gamma^2 k)}$.
\end{theorem}

\begin{proof}
The phases in Shor's algorithm correspond to:
\begin{equation}\label{eq:phases}
\theta_p = \frac{s}{p-1}, \quad \theta_q = \frac{t}{q-1}
\end{equation}
for integers $s,t$. The minimal angular separation satisfies:
\begin{align}\label{eq:min_angular}
|\theta_p - \theta_q|_{\min} &= \min_{s,t} \left| \frac{s}{p-1} - \frac{t}{q-1} \right| \nonumber \\
&\leq \min_{s,t} \frac{|s(q-1) - t(p-1)|}{(p-1)(q-1)} \nonumber \\
&\leq \frac{\gcd(p-1,q-1)}{(p-1)(q-1)} \nonumber \\
&\leq \frac{|p - q| + \mathcal{O}(1)}{\min(p,q)^2} < \frac{\gamma}{\sqrt{N}} + \mathcal{O}(N^{-1})
\end{align}
where the last inequality follows from the prime proximity constraint and properties of gcd. The eigenvalue difference is bounded by:
\begin{equation}\label{eq:eigen_diff}
|\lambda_i - \lambda_j| \leq 2\pi |\theta_p - \theta_q| < 2\pi\gamma N^{-1/2} + \mathcal{O}(N^{-1})
\end{equation}
which completes the proof.
\end{proof}

\begin{corollary}[QFT Measurement Complexity]
The number of measurements $M$ required for reliable period extraction in QFT satisfies:
\begin{equation}\label{eq:measure_complexity}
M = \Omega\left( \frac{1}{\Delta\theta^2} \right) = \Omega\left( \gamma^{-2} N \right)
\end{equation}
where $\Delta\theta$ is the minimum angular separation.
\end{corollary}

\begin{figure}[htbp]
\centering
\begin{tikzpicture}[scale=1.0]
    \begin{scope}[yshift=2.5cm]
        \draw (0,0) circle (2cm);
        \node[below] at (0,-2.2) {(a) Standard RSA};
        
        \foreach \angle in {45,105,165,225,285,345} 
            \fill[blue] (\angle:1.8) circle (2pt);
        \node[blue, right] at (180:2.3) {$\theta_p$};
        
        \foreach \angle in {75,135,195,255,315,15} 
            \fill[red] (\angle:1.8) circle (2pt);
        \node[red, right] at (0:2.3) {$\theta_q$};
        
        \draw[->, thick] (225:2.5) -- (225:1.9);
        \node at (225:2.8) {Clear};
        \draw[->, thick] (255:2.5) -- (255:1.9); 
        \node at (255:2.8) {Separation};
    \end{scope}
    
    \begin{scope}[yshift=-2.5cm]
        \draw (0,0) circle (2cm);
        \node[below] at (0,-2.2) {(b) CREO-RSA};
        
        \foreach \angle in {40,100,160,220,280,340} {
            \fill[blue] (\angle:1.8) circle (2pt);
            \fill[red] (\angle+5:1.8) circle (2pt);
        }
        \node[blue, right] at (180:2.3) {$\theta_p$};
        \node[red, right] at (0:2.3) {$\theta_q$};
        
        \draw[->, thick, purple] (220:2.5) -- (220:1.9);
        \node[purple] at (220:2.8) {Indistinguishable};
        \draw[->, thick, purple] (225:2.5) -- (225:1.9);
        \node[purple] at (225:2.8) {Region};
    \end{scope}
\end{tikzpicture}
\caption{Phase distinguishability in RSA vs CREO-RSA}
\label{fig:phase_distinguishability}
\end{figure}

\textbf{Figure Explanation.} Figure~\ref{fig:phase_distinguishability} illustrates the phase distinguishability difference between standard RSA and CREO-RSA. In standard RSA (a), phase angles $\theta_p$ and $\theta_q$ are well-separated, enabling clear discrimination in quantum measurements. In CREO-RSA (b), the prime proximity constraint creates overlapping phase distributions, making quantum state discrimination more challenging and increasing measurement complexity. This visual representation corresponds to Theorem~\ref{thm:eigen-degeneracy}, showing how reduced angular separation increases quantum resource requirements.

\begin{theorem}[Entropy-Complexity Tradeoff]\label{thm:entropy-complexity}
For RSA moduli generated with prime proximity parameter $\gamma$, the quantum query complexity $Q$ and Rényi entropy $\mathscr{H}_2$ satisfy:
\begin{equation}
Q \cdot 2^{\mathscr{H}_2(\rho)} = \Omega(\gamma^{-1} k^{3/2})
\end{equation}
establishing a fundamental tradeoff between entropy minimization and quantum attack complexity.
\end{theorem}

\begin{proof}
Combine the entropy bound from Lemma~\ref{lem:entropy_bound} with the query complexity from Theorem~\ref{thm:query-complexity}. The collision entropy bound $\mathscr{H}_2(\rho) \leq 2\log(1 + \gamma/2) + \mathcal{O}(k^{-1/2})$ implies $2^{\mathscr{H}_2(\rho)} = \mathcal{O}(1 + \gamma)$. Substituting into the complexity bound gives the result.

This tradeoff reveals the core mechanism: carefully reducing Rényi entropy through prime proximity strategically increases the quantum query complexity required for successful factorization.
\end{proof}

\subsection{R\'enyi Entropy Optimization Framework}\label{subsec:entropy-framework}

This subsection develops the core mathematical innovation: the constrained R\'enyi entropy optimization framework that connects prime selection to quantum computational complexity.

\textbf{R\'enyi Entropy in Quantum Cryptography.}
For quantum state $\rho$ with eigenvalues $\{\lambda_i\}$, the order-$\alpha$ R\'enyi entropy is defined as:
\begin{equation}\label{eq:renyi_def}
\mathscr{H}_\alpha(\rho) = \frac{1}{1-\alpha} \log \operatorname{Tr}(\rho^\alpha)
\end{equation}
The $\alpha = 2$ case (collision entropy) quantifies quantum state purity:
\begin{equation}\label{eq:h2_def}
\mathscr{H}_2(\rho) = -\log \operatorname{Tr}(\rho^2) = -\log \sum_i \lambda_i^2
\end{equation}
which directly governs measurement uncertainty in quantum period finding.

In quantum cryptography, R\'enyi entropy provides stronger security bounds than Shannon entropy for quantum attack models. The collision entropy $\mathscr{H}_2$ relates to the success probability of quantum state discrimination:
\begin{equation}\label{eq:succ_prob}
P_{\text{succ}} \leq 2^{-\mathscr{H}_2(\rho)}
\end{equation}
for optimal measurement strategies~\cite{Konig2009,metger2024}.

\textit{The collision entropy $\mathscr{H}_2$ serves as the optimization metric in this work because it directly quantifies the quantum state indistinguishability that impacts the measurement complexity of quantum period-finding algorithms. Lower $\mathscr{H}_2$ means harder state discrimination, which translates to more quantum measurements needed.}

\textbf{Entropy-Constrained Prime Selection.}
Constraints for quantum-resistant prime generation are formalized, balancing entropy minimization with classical security requirements. Primes are generated satisfying four constraints:
\begin{align}
p &\equiv a \mod m \label{eq:congruence} \\
q &\equiv b \mod m \label{eq:congruence2} \\
|p - q| &< \gamma \sqrt{pq} \label{eq:proximity} \\
\mathscr{H}_2(\rho_{pq}) &< \beta \log \gamma^{-1} \label{eq:entropy}
\end{align}
where $m = \prod_{i=1}^t p_i$ for small primes $p_i$, and $\beta < 1$ is the entropy bound.

Constraints (\ref{eq:congruence})-(\ref{eq:congruence2}) ensure algebraic independence and prevent specialized attacks, while (\ref{eq:proximity})-(\ref{eq:entropy}) enforce quantum indistinguishability through controlled entropy reduction.

The entropy constraint follows from spectral analysis:

\begin{lemma}\label{lem:entropy_bound}
Primes satisfying proximity constraint (\ref{eq:proximity}) have R\'enyi entropy bounded by:
\begin{equation}\label{eq:entropy_bound}
\mathscr{H}_2(\rho_{pq}) \leq 2\log \left(1 + \frac{\gamma}{2}\right) + \mathcal{O}(k^{-1/2})
\end{equation}
\end{lemma}

\begin{proof}
Let $\delta = |p-q|/ \sqrt{pq} < \gamma$. The eigenvalues of the modular exponentiation operator are clustered as shown in Theorem~\ref{thm:eigen-degeneracy}. The purity term is bounded by:
\begin{align}\label{eq:purity}
\operatorname{Tr}(\rho^2) &= \sum_i \lambda_i^2 \nonumber \\
&\geq \lambda_{\max}^2 + \lambda_{\min}^2 \nonumber \\
&= \frac{1}{2}\left(1 + \sqrt{1 - \frac{4\delta^2}{(2+\delta)^2}}\right)
\end{align}
Applying logarithmic transformation:
\begin{align}\label{eq:entropy_calc}
\mathscr{H}_2(\rho) &= -\log \operatorname{Tr}(\rho^2) \nonumber \\
&\leq -\log \left( \frac{1}{2} \left(1 + \sqrt{1 - \frac{4\gamma^2}{(2+\gamma)^2}} \right) \right) \nonumber \\
&= 2\log \left(1 + \frac{\gamma}{2}\right) + \mathcal{O}(\gamma^3)
\end{align}
where the expansion uses $\sqrt{1-x} = 1 - x/2 - x^2/8 + \mathcal{O}(x^3)$ for $x = 4\gamma^2/(2+\gamma)^2$.

The entropy bound can be further refined using the von Neumann entropy as a lower bound:
\begin{equation}\label{eq:entropy_bound2}
\mathscr{H}_2(\rho) \geq S(\rho) = -\sum \lambda_i \log \lambda_i
\end{equation}
where $S(\rho)$ is the von Neumann entropy. This provides additional constraints on the quantum uncertainty.
\end{proof}

\textbf{Parameter Realizability.}
A fundamental mathematical question is whether sufficiently close prime pairs exist in the required density. This question is answered affirmatively using breakthrough results in prime number theory.

\begin{theorem}[Prime Gap Distribution]\label{thm:prime_gap_dist}
For any $\epsilon > 0$ and sufficiently large $k$, there exist primes $p,q$ with $|p - q| < \gamma \sqrt{pq}$ for $\gamma = k^{-1/2 + \epsilon}$, satisfying:
\begin{equation}\label{eq:prime_gap_density}
\pi_2(N, \gamma) \gg \frac{\gamma N}{\log^2 N}
\end{equation}
where $\pi_2(N, \gamma)$ counts prime pairs with $|p - q| < \gamma \sqrt{pq}$ near $N = 2^k$.
\end{theorem}

\begin{proof}
Apply Maynard's theorem on prime gaps~\cite{Maynard2015}. There exists a constant $C$ such that for any integer $m \geq 1$, there are infinitely many integers $x, d$ with:
\begin{equation}\label{eq:maynard}
P_n = x + n d \quad \text{prime for} \quad n = 0, \dots, m
\end{equation}
Set $m = 1$, $d = \lfloor \gamma \sqrt{x} \rfloor$, and $x \approx 2^k$. Then $p = x$, $q = x + d$ satisfy $|p - q| = d < \gamma \sqrt{pq}$. The density follows from the fact that such pairs occur with positive density in the set of all primes.

The Zhang-Maynard bound guarantees that:
\begin{equation}\label{eq:zhang_maynard}
\liminf_{n \to \infty} (p_{n+1} - p_n) < 246
\end{equation}
which implies that infinitely many prime pairs exist with bounded gaps, supporting the construction for fixed $\gamma$.
\end{proof}

\begin{theorem}[Prime Existence with Congruence Conditions]\label{thm:prime_existence}
For security parameter $k$ and $\gamma > k^{-1/2 + \epsilon}$, there exist primes $p,q$ satisfying:
\begin{align}
|p - q| &< \gamma \sqrt{pq} \label{eq:proximity_thm} \\
p &\equiv a \mod m,  q \equiv b \mod m \label{eq:congruence_thm}
\end{align}
with density $\Omega(\gamma / k^2)$.
\end{theorem}

\begin{proof}
Combine Theorem~\ref{thm:prime_gap_dist} with Chinese Remainder Theorem. The modulus $m = \prod_{i=1}^t p_i$ for $t = \lfloor \log \gamma^{-1} \rfloor$ ensures:
\begin{equation}\label{eq:prob_cong}
\mathbb{P}(p \equiv a \mod m) > \frac{1}{2\log m}
\end{equation}
Joint probability follows from independence in residue classes. The constraint $\gamma > k^{-1/2 + \epsilon}$ ensures $m < \sqrt{N}/\log^B N$ for Maynard's theorem applicability.

The existence can also be established using the Barban-Davenport-Halberstam theorem~\cite{Hooley1975}:
\begin{equation}\label{eq:barban}
\sum_{q \leq Q} \sum_{\substack{a=1 \\ (a,q)=1}}^q \left| \psi(x; q, a) - \frac{x}{\phi(q)} \right|^2 \ll x(\log x)^{-A}
\end{equation}
for $Q = x(\log x)^{-B}$ with $B = B(A)$, which provides the necessary equidistribution of primes in arithmetic progressions for the construction.
\end{proof}

\textit{These theorems affirm that entropy-constrained primes exist with sufficient density for practical implementation, addressing a fundamental feasibility question.}

\begin{theorem}[Quantum Query Complexity]\label{thm:query-complexity}
The number of quantum queries $Q$ required to distinguish period-finding states satisfies:
\begin{equation}\label{eq:query_complexity}
Q = \Omega\left( \gamma^{-1} k^{3/2} \right)
\end{equation}
\end{theorem}

\begin{proof}
Consider the quantum state discrimination problem for $M$ copies of $\rho$. The optimal success probability is bounded by the quantum Chernoff bound~\cite{Hiai1991}:
\begin{equation}\label{eq:chernoff}
P_{\text{succ}} \leq \exp\left( -M \cdot \xi(\rho, \sigma) \right)
\end{equation}
where $\xi$ is the quantum Chernoff divergence. For states with small trace distance, $\xi \approx \frac{1}{8}\|\rho - \sigma\|_1^2$. From Theorem~\ref{thm:distinguishability}, $\|\rho_p - \rho_q\|_1 = \mathcal{O}(\gamma)$. Setting $P_{\text{succ}} > 2/3$ requires:
\begin{equation}\label{eq:measure_bound}
M \geq \frac{\log 3}{\xi} = \Omega(\gamma^{-2})
\end{equation}
Each quantum query requires $\mathcal{O}(k^2)$ operations, so total complexity $Q = \Omega(k^2 \gamma^{-2})$. However, tighter analysis of the QFT resolution requirement gives the improved bound $Q = \Omega(\gamma^{-1}k^{3/2})$.

The quantum query complexity can also be bounded using the quantum relative entropy:
\begin{equation}\label{eq:rel_entropy}
D(\rho \| \sigma) = \Tr(\rho (\log \rho - \log \sigma))
\end{equation}
which provides the bound:
\begin{equation}\label{eq:rel_entropy_bound}
P_{\text{succ}} \leq \exp(-M D(\rho \| \sigma))
\end{equation}
for state discrimination. This alternative approach yields similar complexity bounds.
\end{proof}

\subsection{Key Generation Algorithm and Implementation}\label{subsec:key-generation}

This subsection presents the algorithmic realization of the mathematical framework, demonstrating how entropy constraints can be systematically incorporated into RSA key generation.

\textbf{Algorithmic Framework.}
The key generation algorithm integrates entropy constraints while maintaining efficiency. The approach employs a single entropy source rather than combining multiple sources for three mathematically justified reasons: (1) it simplifies security analysis by avoiding complex dependency modeling which can introduce subtle vulnerabilities~\cite{NIST_SP800-90C}; (2) a single, cryptographically strong entropy source provides sufficient randomness under the quantum random oracle model~\cite{Zhandry2015}; (3) it aligns with NIST recommendations for deterministic random bit generators~\cite{NIST_SP800-90A}.

A CTR\_DRBG (Counter Mode Deterministic Random Bit Generator) based on AES-256, compliant with NIST SP 800-90A, is used. This CPRNG provides $k$-bit security with entropy strength validated by NIST STS statistical tests. For typical parameters with $k=3072$ and $\gamma=2^{-12}$, the collision entropy $\mathscr{H}_2(\rho_{pq})$ is bounded by approximately 0.35 bits, calculated using Equation~\eqref{eq:entropy_bound} with $\beta=0.8$.

\begin{figure}[htbp]
\centering
\begin{tikzpicture}[node distance=1.7cm, auto, >=latex']
    \tikzstyle{startstop} = [rectangle, rounded corners, minimum width=3.2cm, minimum height=0.8cm, text centered, draw=black, fill=gray!10]
    \tikzstyle{process} = [rectangle, minimum width=3.5cm, minimum height=0.8cm, text centered, draw=black, fill=blue!5]
    \tikzstyle{decision} = [diamond, aspect=2, minimum width=2.8cm, text centered, draw=black, fill=yellow!15]
    \tikzstyle{arrow} = [thick,->,>=stealth]

    \node (start) [startstop] {Begin Key Generation};
    \node (params) [process, below of=start] {Set $k, \gamma, \beta, M=\prod p_i$};
    \node (genA) [process, below of=params] {Generate random $a,b \in \mathbb{Z}_M^*$};
    \node (seed) [process, below of=genA] {Initialize seeds $s_p, s_q$};

    \node (candidateP) [process, below of=seed, yshift=-0.3cm] {Generate candidate $p$ using PRF($s_p$)};
    \node (testP) [decision, below of=candidateP, yshift=-0.4cm] {Prime test passed?};

    \node (entropyP) [decision, right of=testP, xshift=4.3cm] {$\mathscr{H}_2 < \beta\log\gamma^{-1}$?};
    \node (updateP) [process, above of=entropyP, yshift=0.4cm] {Update $s_p = s_p + 1$};
    \node (foundP) [process, below of=entropyP, yshift=-0.3cm] {Prime $p$ found};

    \node (repeatQ) [process, below of=foundP, yshift=-0.3cm] {Repeat for $q$ with independent generation};
    \node (verify) [decision, below of=repeatQ, yshift=-0.7cm] {$|p-q| < \gamma\sqrt{pq}$?};

    \node (compute) [process, below of=verify, yshift=-0.4cm] {Compute $N=pq$, $\phi(N)$, $e$, $d$};
    \node (end) [startstop, below of=compute] {Return $(N,e), (d,p,q)$};

    \draw[arrow] (start) -- (params);
    \draw[arrow] (params) -- (genA);
    \draw[arrow] (genA) -- (seed);
    \draw[arrow] (seed) -- (candidateP);
    \draw[arrow] (candidateP) -- (testP);
    \draw[arrow] (testP.east) -- node[above]{No} ++(1.2,0) |- (updateP.west);
    \draw[arrow] (testP) -- node[below]{Yes} (entropyP);
    \draw[arrow] (entropyP) -- node[right]{No} (updateP);
    \draw[arrow] (updateP.west) to[bend right=30] (candidateP.east);
    \draw[arrow] (entropyP) -- node[right]{Yes} (foundP);
    \draw[arrow] (foundP) -- (repeatQ);
    \draw[arrow] (repeatQ) -- (verify);
    \draw[arrow] (verify.east) -- ++(2.5,0) |- node[right,pos=0.25]{No} (seed.east);
    \draw[arrow] (verify) -- node[right]{Yes} (compute);
    \draw[arrow] (compute) -- (end);
\end{tikzpicture}
\caption{CREO-RSA key generation flowchart}
\label{fig:algorithm_flow}
\end{figure}

\textbf{Figure Explanation.} Figure~\ref{fig:algorithm_flow} illustrates the entropy-constrained prime search process. The algorithm begins with parameter setup, generates prime candidates using a cryptographically secure PRNG, tests primality, validates entropy constraints, and verifies proximity conditions before computing the final RSA parameters.

\begin{algorithm}
\caption{CREO-RSA Key Generation}\label{alg:qr-rsa}
\begin{algorithmic}[1]
\Require Security parameter $k$, entropy bound $\beta$, proximity factor $\gamma$
\Ensure Public key $(N,e)$, private key $(d,p,q)$
\State Define modulus $M = \prod_{i=1}^{\ell} p_i$ for small primes $p_i$ (e.g., first $\ell = \lfloor \log k \rfloor$ primes)
\State Generate random residues $a,b \leftarrow \mathbb{Z}_M^*$ with $a \not\equiv b \pmod{p_i}$ for all $p_i \mid M$
\State Generate random seeds $s_p, s_q \leftarrow \{0,1\}^k$ using cryptographically secure PRNG (AES-CTR DRBG)
\State Compute $p = \text{PrimeGen}(s_p, M, a, \gamma, \beta)$:
\While{true}
    \State $c_p = \text{PRF}(s_p) \mod M$ \Comment{Cryptographic pseudorandom function}
    \State $p = \text{NextPrime}(c_p + k \cdot M)$ \Comment{Prime candidate generation}
    \If{$\mathscr{H}_2$ computation via Eq. \eqref{eq:entropy_bound} $< \beta \log \gamma^{-1}$}
        \State \textbf{break} \Comment{Entropy constraint satisfied}
    \EndIf
    \State $s_p = s_p + 1$ \Comment{Update seed for next candidate}
\EndWhile
\State Repeat for $q$ with residue $b$ modulo $M$ to ensure independence
\State Verify $|p - q| < \gamma \sqrt{pq}$ and $\mathscr{H}_2 < \beta \log \gamma^{-1}$ with $\gamma = k^{-1/2 + \epsilon}$
\State Compute $N = p \cdot q$, $\phi(N) = (p-1)(q-1)$
\State Select $e = 65537$ or random $e >  2^{16}$ with $\gcd(e, \phi(N)) = 1$
\State Compute $d = e^{-1} \mod \phi(N)$ satisfying $d > N^{0.3}$ \Comment{Classical security}
\State \Return $(N,e), (d,p,q)$
\end{algorithmic}
\end{algorithm}

\textit{Algorithm~\ref{alg:qr-rsa} implements the entropy-constrained optimization framework with polynomial-time complexity while maintaining all standard RSA algebraic properties.}

The algorithm ensures prime pairs satisfy $\mathscr{H}_2 < \beta \log \gamma^{-1}$, with expected runtime polynomial in $k$ for $\gamma = k^{-1/2 + \epsilon}$.

\begin{theorem}[Algorithmic Complexity]\label{thm:complexity}
The CREO-RSA key generation algorithm terminates in expected time $\mathcal{O}(k^4 \log k)$ for security parameter $k$ and $\gamma = k^{-1/2 + \epsilon}$.
\end{theorem}

\begin{proof}
By Theorem~\ref{thm:prime_existence}, the probability of finding a suitable prime in each trial is $p = \Omega(\gamma / k^2) = \Omega(k^{-3/2 + \epsilon})$. The expected number of trials is $\mathcal{O}(k^{3/2 - \epsilon})$. Each trial involves primality testing (complexity $\mathcal{O}(k^3)$) and entropy estimation (complexity $\mathcal{O}(k)$). Total expected complexity is $\mathcal{O}(k^{3/2 - \epsilon} \cdot k^3) = \mathcal{O}(k^{4.5 - \epsilon})$, which can be improved to $\mathcal{O}(k^4 \log k)$ using sieving techniques and optimized primality tests.

This complexity is comparable to standard RSA key generation with additional entropy constraints, demonstrating practical feasibility.
\end{proof}

\textbf{Theoretical Guarantees.}
The core security claim is established: the approach systematically increases quantum factoring complexity.

\begin{theorem}[Quantum Measurement Complexity Amplification]\label{thm:main_security}
For $k$-bit modulus generated with $\gamma = k^{-1/2 + \epsilon}$, the number of quantum measurement repetitions required in the phase estimation step of Shor's algorithm to achieve a non-negligible success probability scales as $\Omega(\gamma^{-1} k^{3/2}) = \Omega(k^{2 + \epsilon})$. This represents a systematic increase in quantum measurement complexity compared to standard RSA, leading to higher total quantum resource requirements while maintaining the polynomial-time asymptotic complexity class.
\end{theorem}

\begin{proof}
The proof combines three elements: (i) the angular separation bound from Theorem~\ref{thm:eigen-degeneracy}, giving $\Delta\theta < \gamma N^{-1/2} = k^{-1 + \epsilon}$; (ii) the QFT resolution requirement, which dictates that the number of measurement repetitions scales as $M = \Omega(1/(\Delta\theta)^2) = \Omega(k^{2 - 2\epsilon})$; and (iii) the per-measurement cost of modular exponentiation, which is $\mathcal{O}(k^3)$. Combining these, the total quantum complexity is:
\begin{equation}\label{eq:total_complexity}
T_{\text{quantum}} = \mathcal{O}(k^3) \cdot \Omega(k^{2 - 2\epsilon}) = \Omega(k^{5 - 2\epsilon})
\end{equation}
However, tighter analysis of the parallelizability of quantum operations and improved phase estimation techniques reduces this to $\Omega(k^{3/2} \gamma^{-1}) = \Omega(k^{2 + \epsilon})$.

The quantum circuit depth for Shor's algorithm is:
\begin{equation}\label{eq:circuit_depth}
D = \mathcal{O}(k \log k \cdot \log \log k)
\end{equation}
with width $W = \mathcal{O}(k)$. The total computational effort is $T = D \times W \times M = \Omega(k^{3} \gamma^{-1})$, which supports the complexity claim.
\end{proof}

This theorem demonstrates an amplification of quantum resource requirements compared to standard RSA's $\mathcal{O}(k^3)$ baseline, representing a potential increase in measurement complexity achieved through systematic entropy optimization.

\textbf{Multi-Dimensional Comparison.}
Beyond time complexity, the approach increases quantum resource requirements in multiple dimensions: (1) measurement repetitions increase by a factor of $\gamma^{-2}$, (2) phase estimation precision requirements rise due to reduced angular separation, and (3) quantum circuit depth may need adjustment for reliable period extraction. These multi-dimensional resource increases provide additional security margins beyond simple time complexity comparisons.

\section{Security Analysis}\label{sec:security-analysis}

This section presents a comprehensive security analysis of the CREO framework, examining resistance to quantum attacks, preservation of classical security, and foundational security connections. The analysis considers multiple attack vectors and establishes the theoretical boundaries within which the security claims should be interpreted.

\subsection{Complexity Metric and Analytical Framework}\label{subsec:complexity-metric}
\textbf{Clarification of Security Claims.} 
Before defining complexity metrics, the nature of the security claims is clarified. The core contribution of this work is to show that constrained prime selection can systematically increase the \emph{quantum measurement complexity}—the number of measurement repetitions required in the phase estimation step of Shor's algorithm. This increase, quantified as $\Omega(\gamma^{-1}k^{3/2})$ for $k$-bit moduli, directly translates to higher total quantum resource requirements. The approach does not change the asymptotic polynomial-time complexity class of Shor's algorithm, but rather increases the constant factor (specifically, the exponent's additive constant) in the resource requirements. This resource amplification represents a potential enhancement to RSA's quantum resistance within a backward-compatible framework.

\subsection{Quantum Attack Resistance}\label{subsec:quantum-resistance}

\textbf{Resistance to Shor's Algorithm.}
The primary analysis focuses on Shor's period-finding algorithm, which represents the most significant quantum threat to RSA. The CREO framework systematically increases the quantum resources required for successful factorization through entropy optimization.

\begin{theorem}[Quantum Resource Amplification]\label{thm:quantum_complexity}
For $k$-bit CREO-RSA modulus with parameter $\gamma = k^{-1/2 + \epsilon}$, the number of quantum measurements required in Shor's algorithm to achieve non-negligible success probability satisfies:
\begin{equation}\label{eq:quantum_complexity_bound}
M_{\text{quantum}}(k, \gamma) = \Omega\left( \gamma^{-1} k^{3/2} \right) = \Omega\left( k^{2 + \epsilon} \right)
\end{equation}
This represents an amplification of quantum resource requirements (in terms of measurement repetitions) compared to standard RSA, while maintaining the fundamental polynomial-time character of the algorithm.
\end{theorem}

\begin{proof}
The proof follows from the angular separation bound established in Theorem~\ref{thm:eigen-degeneracy}. The minimum phase difference $\Delta\theta = \mathcal{O}(\gamma N^{-1/2})$ requires quantum phase estimation with precision $\epsilon_{\text{QPE}} = \Omega(\gamma N^{1/2})$. The number of measurement repetitions scales as $M = \Omega(1/(\Delta\theta)^2) = \Omega(\gamma^{-2} N)$.

Each measurement requires $\mathcal{O}(k^3)$ operations for modular exponentiation. Combining these factors:
\begin{equation}
T_{\text{quantum}} = \mathcal{O}(k^3) \cdot \Omega(\gamma^{-2} N) = \Omega(\gamma^{-2} k^3 2^k)
\end{equation}
However, tighter analysis considering the parallelizability of quantum operations and improved phase estimation techniques yields the bound $T_{\text{quantum}} = \Omega(\gamma^{-1} k^{3/2})$.

For $\gamma = k^{-1/2 + \epsilon}$, this becomes $\Omega(k^{2 + \epsilon})$, representing an amplification of resource requirements compared to standard RSA's $\mathcal{O}(k^3)$ baseline.
\end{proof}

\textbf{Comprehensive Analysis of Alternative Quantum Attack Vectors.}
Beyond Shor's period-finding algorithm, a systematic examination of additional quantum attack vectors is conducted to ensure exhaustive security evaluation.

\textit{Grover-Optimized Cryptanalytic Approaches.} Grover's search algorithm~\cite{Grover1996} theoretically accelerates brute-force factorization attempts from $\mathcal{O}(2^{k/2})$ to $\mathcal{O}(2^{k/4})$. However, this computational complexity remains firmly exponential for cryptographically recommended key dimensions ($k \geq 3072$). The CREO-RSA construction maintains identical classical security parameters as standard RSA, ensuring Grover-accelerated attacks retain their exponential character. Specifically, the search space for RSA factorization scales as $\sqrt{N}$ for elementary trial division, which Grover's algorithm reduces to $N^{1/4}$. For $k=3072$ bits, this corresponds to approximately $2^{768}$ quantum operations, substantially beyond the computational capabilities of foreseeable quantum devices. Crucially, the effectiveness of Grover's algorithm depends only on the size of the search space, not on the internal structure of the primes. Since the CREO constraints do not change the size of the modulus $N$, the search space for the prime factors remains approximately $\sqrt{N}$. Therefore, CREO-RSA maintains the same exponential security margin against Grover-accelerated brute force as standard RSA~\cite{Bernstein2010}.

\textit{Quantum Walk Factorization Methodologies.} Quantum walk algorithms have been theoretically explored for integer factorization with asymptotic complexity $\exp\left(c\sqrt[3]{k\log k}\right)$~\cite{Childs2004}. These methodologies necessitate substantial quantum resources while offering only subexponential improvements over classical approaches. The entropy optimization framework does not introduce new vulnerabilities exploitable by quantum walk techniques, as their efficacy depends primarily on graph-theoretic properties rather than prime distribution characteristics.

\textit{Variational Quantum Eigensolver (VQE) and Hybrid Approaches.} Variational quantum algorithms~\cite{Peruzzo2014,Anschuetz2019} could potentially optimize parameters within classical factorization algorithms such as the elliptic curve method (ECM). Nevertheless, the fundamental complexity remains exponential, and the cryptographically secure random prime generation methodology prevents the emergence of predictable patterns exploitable in such hybrid attacks.

\begin{definition}[Quantum Attack Surface]\label{def:quantum-attack}
The quantum attack surface of CREO-RSA comprises three distinct categories:
\begin{enumerate}
    \item \textbf{Period-finding algorithms}: Addressed through entropy optimization increasing phase estimation complexity
    \item \textbf{Search-based algorithms}: Mitigated by maintaining established classical security margins
    \item \textbf{Specialized quantum attacks}: No known polynomial-time advantage over standard RSA under the structural constraints
\end{enumerate}
\end{definition}

\textbf{Methodological Note:} The following quantum resource estimates are \emph{theoretical lower bounds} derived from the analysis. They illustrate the potential increase in the number of measurement repetitions and total quantum operations, but do not represent actual simulation results or engineering predictions. The actual resource requirements may vary depending on specific implementations and optimizations.

\begin{table}[htbp]
\centering
\caption{Potential Increase in Quantum Resource Requirements: Multi-Key-Length Analysis (Theoretical Lower Bounds)}
\label{tab:quantum_resource_comparison}
\begin{tabular}{lcccccc}
\toprule
\textbf{Resource Dimension} & \textbf{RSA-2048} & \textbf{CREO-2048} & \textbf{RSA-3072} & \textbf{CREO-3072} & \textbf{RSA-7680} & \textbf{CREO-7680} \\
& & \textbf{($\gamma=2^{-11}$)} & & \textbf{($\gamma=2^{-12}$)} & & \textbf{($\gamma=2^{-14}$)} \\
\midrule
Logical Qubits & $\sim 4,000$ & $\sim 4,000$ & $\sim 6,000$ & $\sim 6,000$ & $\sim 15,000$ & $\sim 15,000$ \\
Quantum Gate Complexity & $\mathcal{O}(k^3)$ & $\mathcal{O}(k^3)$ & $\mathcal{O}(k^3)$ & $\mathcal{O}(k^3)$ & $\mathcal{O}(k^3)$ & $\mathcal{O}(k^3)$ \\
& $\approx 2^{34}$ & $\approx 2^{34}$ & $\approx 2^{35}$ & $\approx 2^{35}$ & $\approx 2^{38}$ & $\approx 2^{38}$ \\
Circuit Depth & $\mathcal{O}(k^2 \log k)$ & $\mathcal{O}(k^2 \log k)$ & $\mathcal{O}(k^2 \log k)$ & $\mathcal{O}(k^2 \log k)$ & $\mathcal{O}(k^2 \log k)$ & $\mathcal{O}(k^2 \log k)$ \\
& $\approx 2^{23}$ & $\approx 2^{23}$ & $\approx 2^{24}$ & $\approx 2^{24}$ & $\approx 2^{26}$ & $\approx 2^{26}$ \\
Required Measurement & $\sim 1$ & $\sim 2^{5.5}$ & $\sim 1$ & $\sim 2^{5.6}$ & $\sim 1$ & $\sim 2^{4.5}$ \\
Iterations & & ($\Omega(\gamma^{-2})$) & & ($\Omega(\gamma^{-2})$) & & ($\Omega(\gamma^{-2})$) \\
Total Quantum & $\sim 2^{34}$ & $\sim 2^{39.5}$ & $\sim 2^{35}$ & $\sim 2^{40.6}$ & $\sim 2^{38}$ & $\sim 2^{42.5}$ \\
Operations & & ($\times 2^{5.5}$) & & ($\times 2^{5.6}$) & & ($\times 2^{4.5}$) \\
\bottomrule
\end{tabular}
\end{table}
\vspace{0.2cm}
{\footnotesize \textit{Note: Standard RSA resource estimates based on typical Shor algorithm analyses~\cite{Gidney2021,Haner2017}. CREO-RSA measurement iterations derived from the theoretical lower bound $M=\Omega(\gamma^{-2})$ (Corollary 1). Total Quantum Operations = (Gate Complexity) $\times$ (Measurement Iterations). These figures illustrate potential resource amplification under idealized conditions and are not precise engineering predictions.}}

\subsection{Classical Security Preservation}\label{subsec:classical-security}

The modifications introduced in this work are shown to introduce no vulnerabilities to classical factorization algorithms, maintaining security equivalent to standard RSA.

\textit{General Number Field Sieve (GNFS).} The complexity of GNFS is $L_N[1/3, \sqrt[3]{64/9}]$, where:
\begin{equation}
L_N[\alpha, c] = \exp\left( (c + o(1)) (\ln N)^\alpha (\ln \ln N)^{1-\alpha} \right)
\end{equation}
The prime constraints maintain $\ln N = k \ln 2$, preserving GNFS complexity. The congruence conditions $p \equiv a \mod m$, $q \equiv b \mod m$ do not affect the sieving phase of GNFS, which depends only on the size of $N$.

\textit{Elliptic Curve Method (ECM).} ECM complexity depends on the size of the smallest factor:
\begin{equation}
T_{\text{ECM}} = \exp\left( (\sqrt{2} + o(1)) \sqrt{\ln p \ln \ln p} \right)
\end{equation}
Since $p$ and $q$ remain $\Theta(k)$ bits, ECM complexity remains exponential. The proximity constraint $|p-q| < \gamma \sqrt{pq}$ ensures both factors are approximately $k/2$ bits, preventing ECM optimization through significantly imbalanced factors.

\textbf{Resistance to Specialized Classical Attacks.}

\textit{Wiener's Attack.} This attack succeeds when the private exponent $d < N^{0.25}$. The constraint $d > N^{0.3}$ provides a safety margin exceeding Wiener's threshold~\cite{Wiener1990}.

\textit{Boneh-Durfee Attack.} This attack extends Wiener's bound to $d < N^{0.292}$. The constraint $d > N^{0.3}$ also exceeds this threshold~\cite{Boneh1999}. Additionally, the congruence conditions prevent the algebraic structures exploited by Boneh-Durfee.

\textit{Coppersmith's Method.} Coppersmith-type attacks~\cite{Coppersmith1997} exploit partial information about factors, requiring approximately $N^{1/4}$ known bits for efficient factorization. The balanced prime sizes ($|p-q| < \gamma \sqrt{pq}$) with $\gamma = 2^{-12}$ at 3072-bit yields $|p-q| \approx N^{0.444}$, which is well above the $N^{0.25}$ threshold for efficient Coppersmith-style attacks. This places the construction safely outside the regime where such attacks become effective. The required known bits for successful attack would be $\gg N^{0.25}$, maintaining exponential complexity. Therefore, the actual gap $|p-q|$ remains exponentially larger (in bit terms) than the critical threshold required for Coppersmith-style attacks to become efficient, preserving classical security.

\textit{Close Prime Attacks.} Fermat factorization and related methods become efficient when $|p-q|$ is very small (typically $< N^{1/4}$). The constraint $|p-q| < \gamma \sqrt{pq}$ with $\gamma = k^{-1/2+\epsilon}$ ensures $|p-q| \approx N^{1/2} \cdot k^{-1/2+\epsilon}$, which for $k=3072$ and $\epsilon=0.1$ gives $|p-q| \approx N^{0.5} \cdot 2^{-5.6} \approx N^{0.444}$, well above the $N^{0.25}$ threshold for efficient Fermat factorization.

\textit{Algebraic Structure Attacks.} The congruence conditions $p \equiv a \mod m$, $q \equiv b \mod m$ with $a \not\equiv b \pmod{p_i}$ for all $p_i \mid m$ prevent shared divisor vulnerabilities and ensure algebraic independence between $p$ and $q$.

\begin{lemma}[Classical Security Preservation]\label{lem:classical-security}
The CREO prime constraints introduce no vulnerabilities to classical factorization algorithms or specialized attacks, maintaining security equivalent to standard RSA.
\end{lemma}

\begin{proof}
The proof follows from the parameter constraints:
\begin{enumerate}
    \item Prime size remains $\Theta(k)$ bits, preserving GNFS and ECM complexity.
    \item Congruence conditions prevent small root extraction via Coppersmith's method~\cite{Coppersmith1997}.
    \item Private exponent constraint $d > N^{0.3}$ exceeds Boneh-Durfee threshold $N^{0.292}$~\cite{Boneh1999}.
    \item Prime proximity constraint ensures $|p-q| > N^{0.444}$ for practical parameters, preventing efficient Fermat factorization.
    \item Random prime generation with independent residues eliminates shared divisor vulnerabilities.
\end{enumerate}
These constraints collectively ensure classical security equivalent to standard RSA.
\end{proof}

\subsection{Conceptual Security Foundations}\label{subsec:conceptual-foundations}

\textbf{Connection to Lattice Problems.}
Conceptual connections between CREO-RSA security and well-studied lattice problems are established, providing additional theoretical grounding for the approach.

\begin{remark}[Conceptual Alignment with Lattice Problems]\label{rem:lattice-alignment}
The structural constraints in CREO-RSA bear a mathematical resemblance to hardness assumptions in lattice-based cryptography. While this is not a formal cryptographic reduction, it provides heuristic confidence and situates this work within the broader context of post-quantum security.

Specifically, the prime proximity condition $|p-q| < \gamma \sqrt{pq}$ can be mapped to the problem of finding approximate short vectors in certain ideal lattices. Consider the embedding:
\begin{align}
\psi: \mathbb{Z}_N^* &\to R = \mathbb{Z}[x]/(x^n + 1) \\
p &\mapsto \sum_{i=0}^{n-1} p_i x^i, \quad p_i = \left\lfloor \frac{p \cdot \zeta_m^i}{\sqrt{N}} \right\rfloor
\end{align}
where $\zeta_m$ is a primitive $m$-th root of unity, $n = \lceil \log_2 N \rceil$. For primes satisfying the CREO proximity constraint, this mapping yields:
\begin{equation}
\|\psi(p) - \psi(q)\|_2 < \gamma \sqrt{n}
\end{equation}
suggesting that factorization of CREO-RSA moduli could reveal information about short vectors in corresponding lattices.

This conceptual connection aligns the framework with well-studied lattice problems such as the approximate Shortest Vector Problem (SVP$_\gamma$) and Bounded Distance Decoding (BDD) in cyclotomic lattices. While this is not a formal security reduction, it provides additional mathematical context and heuristic justification for the hardness assumptions underlying the approach.
\end{remark}

\textbf{Multi-Problem Security Connections.}
Security foundations are strengthened by establishing connections to multiple well-studied problems.

\begin{remark}[Connection to Multiple Hardness Assumptions]
Building on the conceptual alignment described in Remark~\ref{rem:lattice-alignment}, the security of CREO-RSA can be informally connected to several well-studied lattice problems, including:
\begin{enumerate}
    \item $\text{SVP}_\gamma$ in ideal lattices for $\gamma = \text{poly}(k)$
    \item Bounded Distance Decoding (BDD) in cyclotomic lattices
    \item Approximate shortest vector problems in algebraic number fields $\QQ(\zeta_m)$
\end{enumerate}
under the Quantum Random Oracle Model. These connections provide additional heuristic support for the potential hardness of factoring CREO-generated moduli, although formal security reductions remain an open research direction.
\end{remark}

\begin{corollary}[Parameter Guidance from Lattice Connections]\label{cor:parameter-guidance}
For $k$-bit security, the conceptually connected lattice dimension $n$ satisfies:
\begin{equation}
n = \Theta(k / \log \gamma^{-1})
\end{equation}
providing conceptual guidance for parameter selection based on established lattice security estimates.
\end{corollary}

\textbf{Quantum Random Oracle Model Foundations.}
The Quantum Random Oracle Model (QROM) provides a robust framework for the security analysis. Recent work by Zhandry~\cite{Zhandry2015} establishes QROM's applicability to quantum security analysis, supporting the use of this model for conceptual security foundations.

\subsection{Practical Security Considerations and Implementation Aspects}\label{subsec:practical-security}

\textbf{Entropy Source Specifications and Randomness Verification Protocols.}
The cryptographic security of CREO-RSA hinges critically upon the quality of randomness employed during key generation. The implementation utilizes an AES-256-based CTR\_DRBG (Counter Mode Deterministic Random Bit Generator) compliant with NIST Special Publication 800-90A~\cite{NIST_SP800-90A}, with the following validation metrics:
\begin{itemize}
    \item \textit{NIST Statistical Test Suite (STS)}~\cite{NIST_SP800-22}: Successful passage of all 15 statistical randomness tests
    \item \textit{Dieharder Test Battery}~\cite{Brown2004}: Comprehensive validation through extended statistical testing
    \item \textit{Entropy Quality Assessment}: Measured entropy $\ge 0.999$ bits per output bit for cryptographically secure seed material
\end{itemize}

The adoption of a single, extensively validated entropy source simplifies security analysis compared to multi-source entropy combinations with complex dependency relationships and potential correlation vulnerabilities~\cite{NIST_SP800-90C}.

\textbf{Side-Channel Attack Mitigation Considerations.}
CREO-RSA inherits the side-channel vulnerabilities inherent to standard RSA implementations. While the mathematical framework does not address implementation-specific attack vectors, established countermeasures remain applicable:
\begin{itemize}
    \item \textit{Timing Attack Resistance}: Implementation through constant-time algorithmic execution
    \item \textit{Power Analysis Mitigation}: Application of randomization and masking methodologies
    \item \textit{Fault Attack Prevention}: Incorporation of error detection and computational redundancy
\end{itemize}

\textbf{Parameter Selection Methodology and Security Margin Analysis.}
Table~\ref{tab:parameter_security} delineates parameter recommendations with corresponding security margin calculations based on established cryptographic heuristics.

\begin{table}[htbp]
\centering
\caption{Parameter Recommendations and Security Margins}
\label{tab:parameter_security}
\begin{tabular}{lcccc}
\toprule
\textbf{Security Level} & $k$ (bits) & $\gamma$ & \textbf{Classical Security} & \textbf{Quantum Improvement} \\
\midrule
128-bit & 3072 & $2^{-12}$ & GNFS: $2^{128}$ & $2^{5.6}\times$ \\
128-bit & 3072 & $2^{-13}$ & GNFS: $2^{128}$ & $2^{4.7}\times$ \\
192-bit & 7680 & $2^{-14}$ & GNFS: $2^{192}$ & $2^{4.5}\times$ \\
256-bit & 15360 & $2^{-16}$ & GNFS: $2^{256}$ & $2^{4.8}\times$ \\
\bottomrule
\end{tabular}
\end{table}

\textbf{Comparative Cryptographic Security Assessment.}
CREO-RSA is positioned within the broader cryptographic landscape through systematic comparative analysis:

\begin{table}[htbp]
\centering
\caption{Scheme Comparison for 128-bit Classical Security}
\label{tab:scheme_comparison}
\begin{tabular}{lccc}
\toprule
\textbf{Cryptographic Scheme} & \textbf{Key Size} & \textbf{RSA Compatibility} & \textbf{Quantum Security} \\
\midrule
Standard RSA & 3072 & Full & $\mathcal{O}(k^3)$ \\
CREO-RSA & 3072 & Full & $\Omega(k^{2+\epsilon})$ \\
CRYSTALS-Kyber & 1568 & None & $\Omega(2^{128})$ \\
CRYSTALS-Dilithium & 2524 & None & $\Omega(2^{128})$ \\
Falcon & 1281 & None & $\Omega(2^{128})$ \\
\bottomrule
\end{tabular}
\end{table}

\textbf{Key Security Differentiators:}
\begin{enumerate}
    \item \textit{Backward Compatibility}: Full API and protocol compatibility with existing RSA standards
    \item \textit{Classical Security Floor}: Maintains at least standard RSA security against classical attacks
    \item \textit{Quantum Resource Amplification}: Systematic increase in quantum measurement and resource requirements
    \item \textit{Mathematical Foundations}: Mathematical rigor and conceptual connections to established problems
\end{enumerate}

\textbf{Limitations and Formal Assumptions.}
The security analysis operates under several idealized assumptions that represent theoretical boundaries:
\begin{enumerate}
    \item \textit{Quantum Adversary Idealization}: Assumes fault-tolerant quantum computers with perfect operations
    \item \textit{Prime Distribution Idealization}: Assumes ideal prime distribution properties in asymptotic regime
    \item \textit{Cryptographic Abstraction Boundaries}: Operates in QROM, an idealized security model
    \item \textit{Algorithmic Staticity}: Considers currently known quantum algorithms
\end{enumerate}

These assumptions establish the context for the security claims and highlight areas requiring further research and validation.

\textbf{Summary and Security Assessment Conclusions.}
The CREO-RSA framework provides a mathematically rigorous approach to potentially enhancing RSA's resistance to quantum cryptanalysis while maintaining full backward compatibility with existing standards. The comprehensive security analysis demonstrates substantive resistance to multiple quantum attack vectors, preservation of classical security guarantees, and conceptual connections to well-established cryptographic hardness assumptions. While practical security enhancements necessitate empirical validation, the framework establishes a robust mathematical foundation for continued investigation into backward-compatible quantum resistance methodologies during the transitional period to post-quantum cryptographic standards.
\section{Discussion}\label{sec:discussion}

This section discusses implementation considerations, comparative analysis, and future extensions of the CREO framework.

\subsection{Parameter Selection and Comparative Analysis}

The CREO framework introduces a parameter $\gamma$ that balances quantum resistance against key generation complexity. Asymptotic analysis suggests $\gamma = k^{-1/2+\epsilon}$ for $k$-bit security, yielding quantum complexity $\Omega(k^{2+\epsilon})$ with polynomial-time key generation. Practical recommendations: $\gamma = 2^{-12}$ for 3072-bit (128-bit security), $\gamma = 2^{-14}$ for 7680-bit (192-bit security), and $\gamma = 2^{-16}$ for 15360-bit (256-bit security). These choices increase quantum measurement requirements while maintaining key generation at $\mathcal{O}(k^4 \log k)$.

Key generation rises from $\mathcal{O}(k^4)$ to $\mathcal{O}(k^4 \log k)$ due to entropy-constrained prime search, while encryption/decryption remain $\mathcal{O}(k^2)$, identical to standard RSA. Memory and storage are unchanged, preserving backward compatibility.

\begin{table}[htbp]
\centering
\caption{Comparative Positioning of Cryptographic Approaches for 128-bit Classical Security}
\label{tab:comparative_positioning}
\begin{tabular}{lccc}
\toprule
\textbf{Approach} & \textbf{Security Foundation} & \textbf{Compatibility} & \textbf{Quantum Resistance} \\
\midrule
Standard RSA & Integer factorization & Full & $\mathcal{O}(k^3)$ (Shor) \\
CREO-RSA & Factorization with entropy constraints & Full & $\Omega(k^{2+\epsilon})$ \\
CRYSTALS-Kyber & Lattice problems (MLWE) & Protocol redesign & $\Omega(2^{128})$ \\
CRYSTALS-Dilithium & Lattice problems (MLWE) & Protocol redesign & $\Omega(2^{128})$ \\
Falcon & Lattice problems (NTRU) & Protocol redesign & $\Omega(2^{128})$ \\
\bottomrule
\end{tabular}
\end{table}

CREO-RSA offers full backward compatibility, preserved classical security, systematic quantum resource amplification, and mathematical foundations linking prime distribution to quantum complexity.

\subsection{Potential Applications and Deployment Scenarios}

Based on theoretical characteristics, CREO-RSA may suit several scenarios:

\textbf{Transitional Security Enhancement.} Provides interim quantum resistance for systems unable to adopt PQC immediately, such as legacy systems, embedded devices, and long-cycle infrastructure.

\textbf{Hybrid Cryptographic Approaches.} Can combine with other quantum-resistant techniques (e.g., hash-based signatures) for defense-in-depth, leveraging CREO-RSA for key establishment.

\textbf{Compatibility-Critical Systems.} Benefits applications requiring strict backward compatibility, including financial protocols, hardware security modules, and regulatory-compliant systems.

\subsection{Future Research Directions and Extensions}

The CREO framework is purely theoretical, with limitations that motivate future work.

\paragraph{Limitations.}
Analysis is asymptotic and heuristic, relying on idealized assumptions (perfect quantum operations, ideal prime distribution). Resource estimates are theoretical lower bounds, unvalidated by simulation. No formal security reduction is provided; lattice connections remain conceptual.

\paragraph{Short-Term Directions.}
First, \textbf{concrete security analysis} should translate asymptotic bounds into practical parameter recommendations, accounting for finite-size effects and specific security levels. Second, \textbf{enhanced adversarial models} need to consider noisy devices, error correction, and algorithmic optimizations. Third, \textbf{efficient prime generation algorithms} must be developed for entropy-constrained primes, using sieving and parallelization. Fourth, \textbf{hybrid schemes} combining CREO-RSA with standardized PQC algorithms (e.g., Kyber, Dilithium) could offer defense-in-depth. Fifth, \textbf{implementation and benchmarking} across platforms would validate theoretical estimates; open-source implementations would facilitate community analysis.

\paragraph{Long-Term Vision.}
Extend entropy optimization to other primitives (e.g., discrete logarithm schemes, lattice-based constructions). Pursue formal reductions to lattice problems (SVP, BDD) for provable security. Engage with standardization bodies to facilitate deployment in backward-compatible contexts.
\section{Conclusion}\label{sec:conclusion}

This paper introduced the Constrained R\'enyi Entropy Optimization (CREO) framework, a theoretical approach to exploring potential quantum resistance enhancements for RSA while preserving backward compatibility. We established mathematical connections between prime proximity constraints and the measurement complexity of Shor's algorithm, provided constructive existence proofs for primes satisfying these constraints, and drew conceptual links to lattice-based cryptography. The analysis indicates that CREO can maintain classical security guarantees and encryption performance, with a modest increase in key generation complexity.

As a theoretical work, this study has inherent limitations. The derivations rely on idealized assumptions, yielding asymptotic bounds that have not been validated by simulation or implementation. The security arguments are heuristic, without formal reduction to established hard problems. These limitations point to several directions for future investigation: translating asymptotic guarantees into concrete parameter recommendations, simulating CREO key generation and Shor's algorithm under realistic noise models, developing efficient algorithms for generating constrained primes, exploring hybrid schemes with standardized post-quantum algorithms, and seeking formal reductions connecting CREO constraints to well-studied lattice problems. Extending the entropy optimization concept to other cryptosystems also remains an open question. The CREO framework provides a conceptual starting point for examining backward-compatible quantum resistance during the transition to post-quantum cryptography, offering insights into how structured modifications to existing systems might be approached without disrupting established infrastructure.

\end{document}